\documentclass[twocolumn]{aastex62}
\usepackage{natbib}
\graphicspath{{./}{figures/}}
\usepackage{color,subfigure,lineno} 

\shorttitle{Variability Biases}
\shortauthors{Shen \& Burke}

\begin{document}

\title{A Sample Bias in Quasar Variability Studies}


\author[0000-0003-1659-7035]{Yue Shen}
\affiliation{Department of Astronomy, University of Illinois at Urbana-Champaign, Urbana, IL 61801, USA}
\affiliation{National Center for Supercomputing Applications, University of Illinois at Urbana-Champaign, Urbana, IL 61801, USA}

\author[0000-0001-9947-6911]{Colin J. Burke}
\affiliation{Department of Astronomy, University of Illinois at Urbana-Champaign, Urbana, IL 61801, USA}
\affiliation{Center for AstroPhysical Surveys, National Center for Supercomputing Applications, University of Illinois at Urbana-Champaign, Urbana, IL 61801, USA}

\begin{abstract}
When a flux-limited quasar sample is observed at later times, there will be more dimmed quasars than brightened ones, due to a selection bias induced at the time of sample selection. Quasars are continuously varying and there are more fainter quasars than brighter ones. At the time of selection, even symmetrical variability will result in more quasars with their instantaneous fluxes scattered above the flux limit than those scattered below, leading to an asymmetry in flux changes over time. The same bias would lead to an asymmetry in the ensemble structure function (SF) of the sample such that the SF based on pairs with increasing fluxes will be slightly smaller than that based on pairs with decreasing fluxes. We use simulated time-symmetric quasar light curves based on the damped random walk prescription to illustrate the effects of this bias. The level of this bias depends on the sample, the threshold of magnitude changes, and the coverage of light curves, but the general behaviors are consistent. In particular, the simulations matched to recent observational studies with decade-long light curves produce an asymmetry in the SF measurements at the few percent level, similar to the observed values. These results provide a cautionary note on the reported time asymmetry in some recent quasar variability studies. 

\end{abstract}


\section{Introduction}\label{sec:introduction}

Recent wide-field optical time-domain surveys have greatly improved the statistics for quasar variability studies \citep[e.g.,][]{VandenBerk_etal_2004,deVries_etal_2005,Bauer_etal_2009,MacLeod_etal_2010,Caplar_etal_2017}. Optical variability of quasars mostly traces the variations from the accretion disk emission, and is observed to be stochastic and ubiquitous over the full ranges of timescales and quasar properties. However, the nature of quasar variability is still largely unknown, and observations of the ensemble quasar variability properties can potentially be used to understand the origin of the optical variability of quasars. For example, \citet{Kawaguchi_etal_1998} proposed to use the SF of quasar variability to distinguish different models, based on the asymmetry in the SF measured from brightening or fading pairs:
\begin{equation}\label{eqn:beta}
\beta(\tau)=\frac{{\rm SF_{ic}(\tau)} - {\rm SF_{dc}(\tau)} }{ {\rm SF_{tot}}(\tau) }\ ,
\end{equation}
where ${\rm SF_{ic}}$, ${\rm SF_{dc}}$, ${\rm SF_{tot}}$ are the SF based on brightening pairs, fading pairs and total pairs, respectively. A negative $\beta$ 
would imply a gradual rise and rapid fading in the light curve, as expected from the model of accretion disk instabilities \citep{Kawaguchi_etal_1998}. While earlier measurements of this $\beta$ statistic were inconclusive based on various quasar samples \citep[e.g.,][]{Giveon_etal_1999,Hawkins_2002,deVries_etal_2005,Bauer_etal_2009,Chen_Wang_2015}, a statistically significant negative $\beta$ was reported in more recent variability studies based on large quasar samples from the SDSS Stripe 82 \citep{Voevodkin_2011} and the CRTS bright quasar sample \citep{Tachibana_etal_2020}, with $\sim 10$ yrs long light curves.

In the meantime, it has gained popularity to model stochastic quasar variability with Gaussian processes. The Damped Random Walk (DRW) model is a simple stationary, time-symmetric Gaussian process shown to fit the general optical variability of quasars well \citep[e.g.,][]{Kelly_etal_2009,MacLeod_etal_2010}. Despite its popularity and overall success in fitting quasar light curves, there are concerns on the applicability of the DRW model \citep[e.g.,][]{Mushotzky_etal_2011}. The reported asymmetry in the SF, if true, would also imply deviations of quasar variability from a DRW, or time-symmetric Gaussian processes in general. 

However, there is a selection bias in the samples of quasars used for long-term variability studies that could potentially lead to artificial trends  misinterpreted as physical. Many recent quasar variability studies are based on spectroscopically confirmed SDSS quasars, with light curves compiled from different photometric surveys later on. The quasar sample defined earlier on is typically flux limited, and contains more fainter (in terms of the mean luminosity, since quasars are variable) quasars scattered above the flux limit than those scattered below, which would eventually scatter down given sufficient time\footnote{In the case of a DRW, the typical turnaround timescale is the damping timescale, which is about hundreds of days in the quasar rest-frame \citep[e.g.,][]{MacLeod_etal_2010}.}. Therefore, statistically the sample will have more dimmed quasars than brightened ones when observed at later epochs, as confirmed in observations \citep[e.g.,][]{Rumbaugh_etal_2018,Caplar_etal_2020}. This asymmetry in quasar brightness changes at later times was interpreted as a selection bias and modeled with simulated light curves \citep[e.g.,][]{Rumbaugh_etal_2018,Luo_etal_2020}. The same bias should also lead to an asymmetry in the SF, with an overall negative $\beta$ as defined in Eqn.~(\ref{eqn:beta}).

In this work we quantify the effects of this sample bias using simulated data. We describe our simulations and main results in \S\ref{sec:method}. We discuss our findings in \S\ref{sec:disc} and conclude in \S\ref{sec:con}. The light curve analysis is performed in magnitude units following the common practice for optical quasar variability studies. 

\section{Methods and Results}\label{sec:method}

\subsection{A toy model}\label{sec:toy}

The aforementioned bias is a common feature of flux-limited samples with scatters in the observed quantities, and is a generalized form of the Eddington bias \citep{Eddington_1913}. Here we use a toy model to illustrate this bias. The instantaneous (noise-free) magnitude $m^\prime$ given mean magnitude $m$ is assumed to be a Gaussian random variable $G(m,\sigma_{\rm lc})$ with mean value $m$ and root-mean-square (RMS) dispersion $\sigma_{\rm lc}$. The underlying quasar luminosity function (LF) $\Phi(m)$ (in units of space density per mag), is assumed to be a single power-law $\Phi(m)\propto 10^{\gamma m}$, where the slope $\gamma>0$ (in units of ${\rm mag}^{-1}$). 
\begin{figure}
  \centering
    \includegraphics[width=0.48\textwidth]{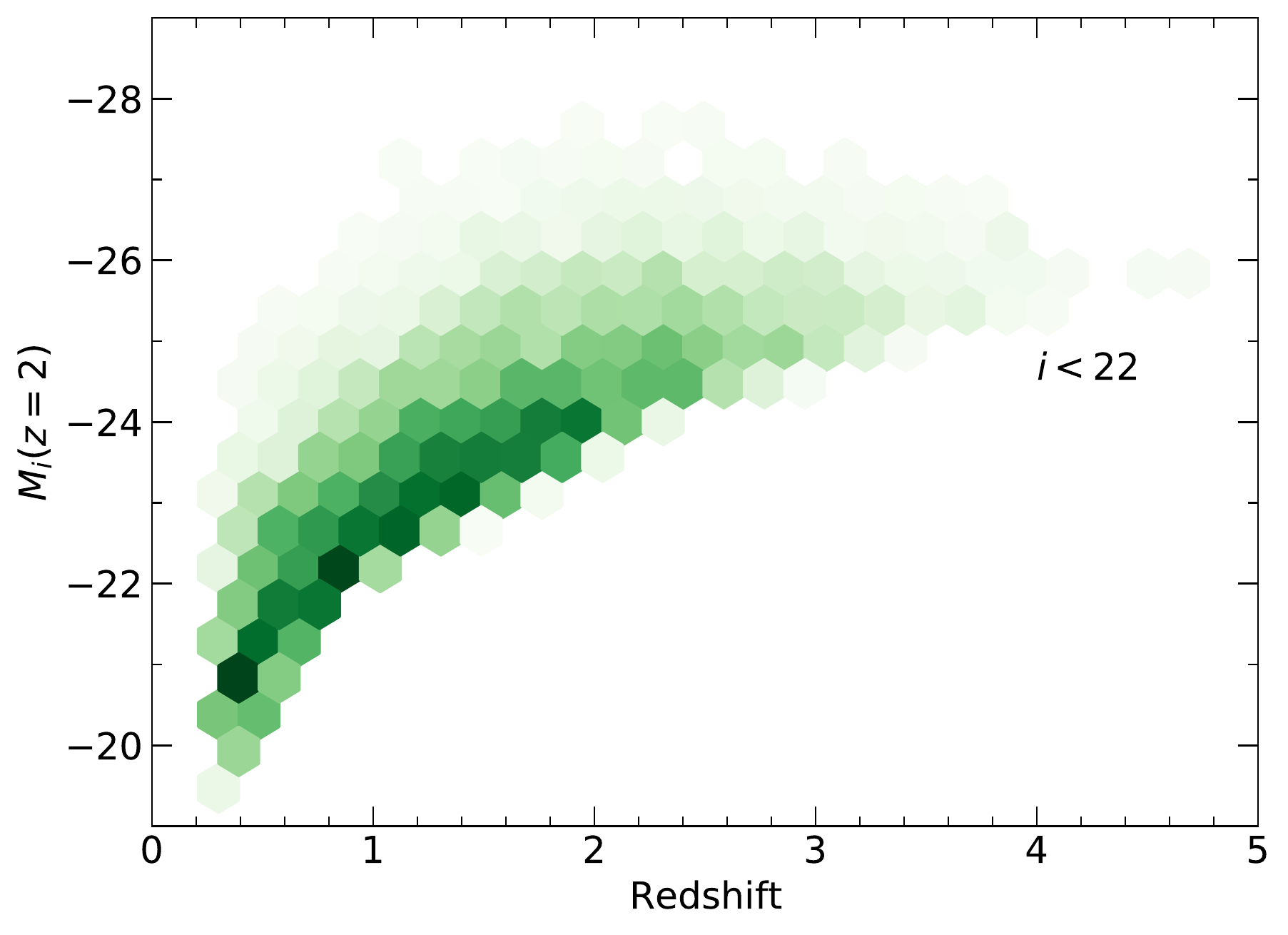}
    \includegraphics[width=0.48\textwidth]{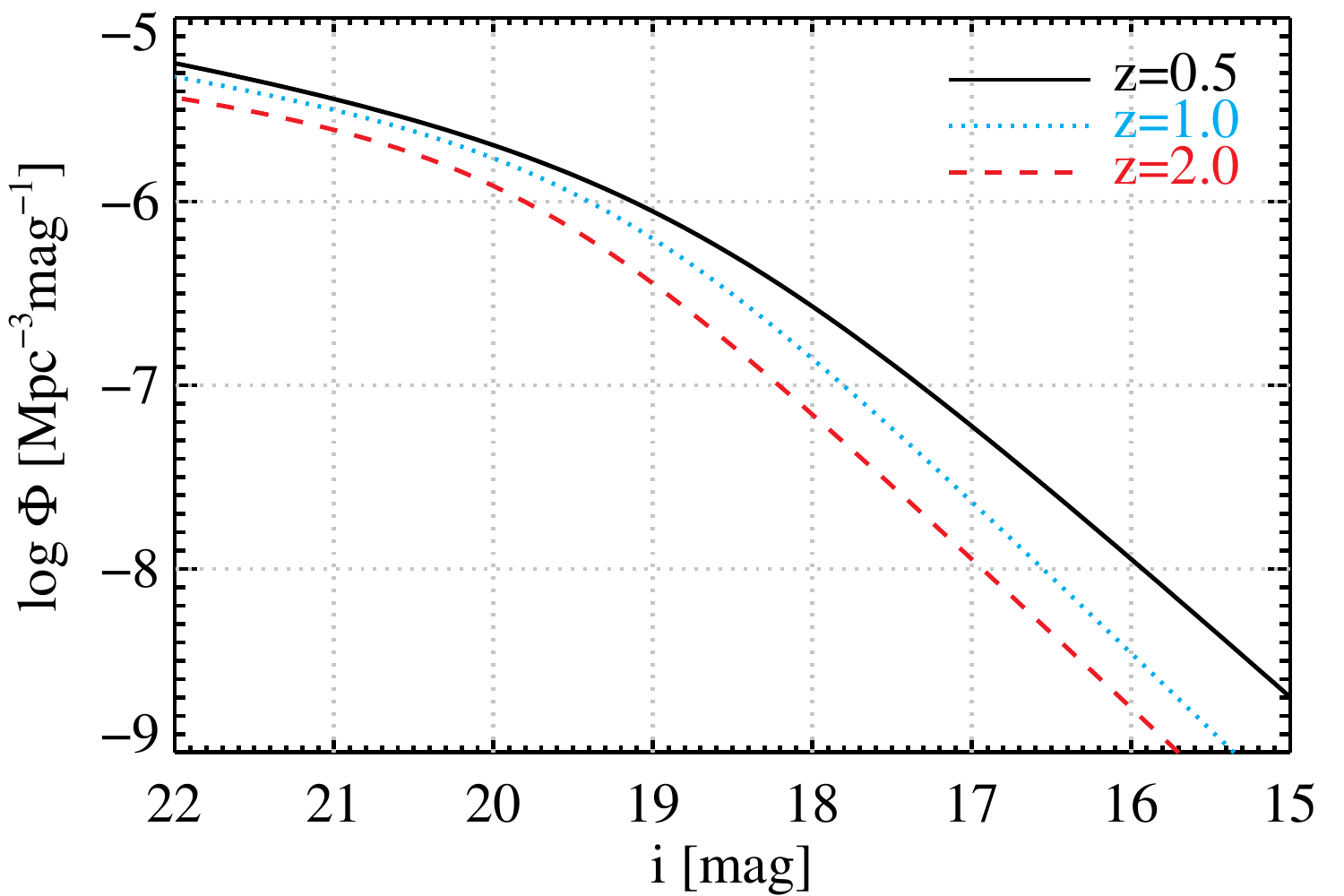}
    \includegraphics[width=0.48\textwidth]{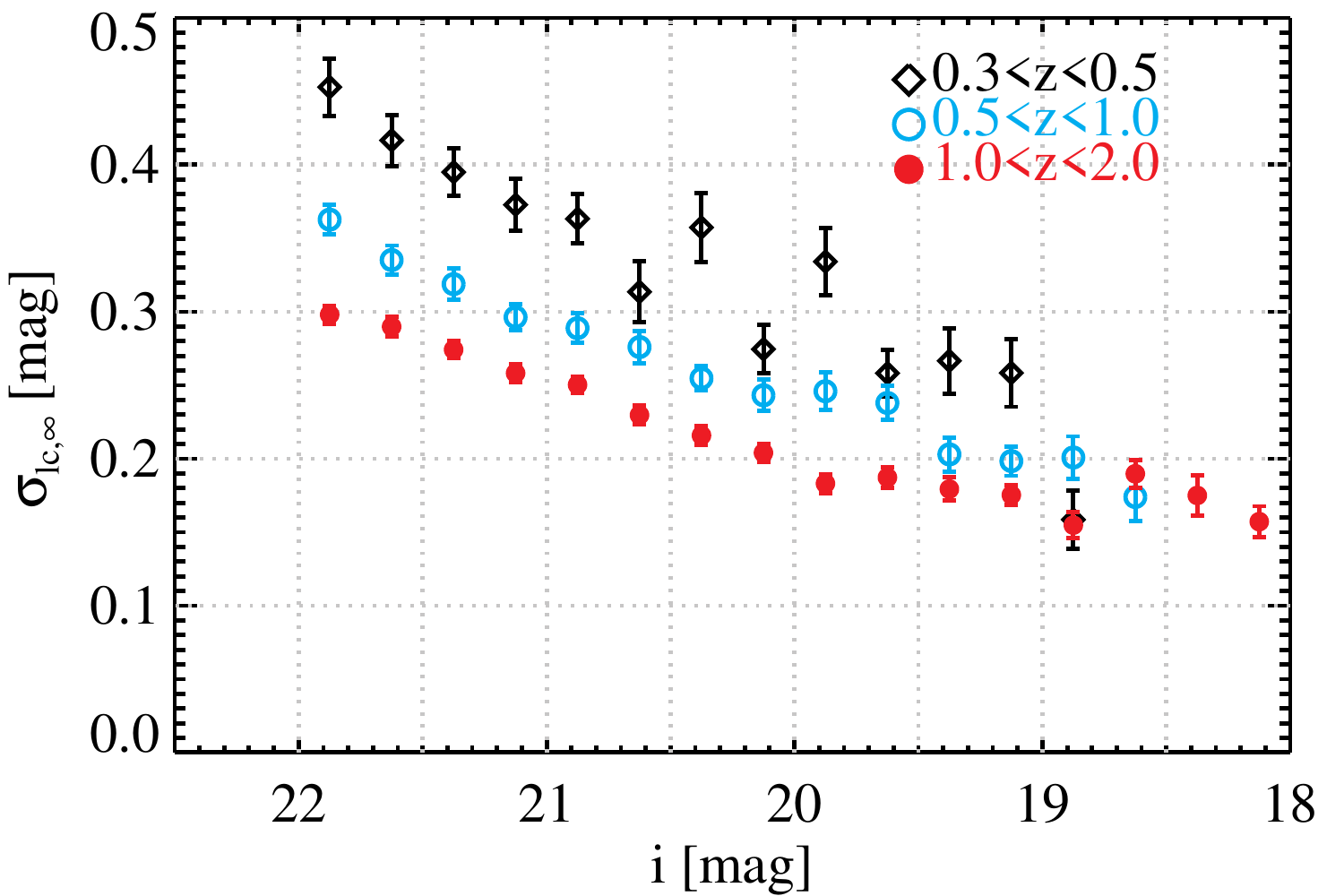}
    \caption{Basic properties of the simulated quasar sample in \S\ref{sec:sims}. {\em Top:} Luminosity \citep[denoted by $M_i(z=2)$, see][]{Richards_etal_2006a} and redshift distribution of the simulated quasar sample (higher density regions denoted by darker colors). {\em Middle:} Adopted LF at three redshifts. {\em Bottom:} Sample-median RMS of the light curves $\sigma_{\rm lc,\infty}={\rm SF_{\infty}}/\sqrt{2}$ in different magnitude-redshift bins. }
    \label{fig:Lz}
\end{figure}

This simple model leads to a constant bias between the instantaneous magnitude and the mean magnitude in the sample \citep[e.g.,][]{Shen_Kelly_2010}:

\begin{equation}\label{eqn:mag_bias}
    A\equiv m^\prime - m = -\ln(10)\gamma \sigma_{\rm lc}^2 \ .
\end{equation}

At each instantaneous magnitude $m^\prime$, the number ratio of quasars with $m>m^\prime+\Delta m$ and with $m<m^\prime-\Delta m$,  where $\Delta m\ge 0$ is a given threshold in magnitude differences, is:
\begin{eqnarray}\label{eqn:toy}
    R&=&\frac{ \int_{m^\prime+\Delta m}^{+\infty} 10^{\gamma m} e^{-\frac{(m - m^\prime)^2}{2\sigma_{\rm lc}^2}}dm }{\int_{-\infty}^{m^\prime-\Delta m} 10^{\gamma m} e^{-\frac{(m - m^\prime)^2}{2\sigma_{\rm lc}^2}}dm }\nonumber\\
    &=&\frac{1 - {\rm erf}\bigg[(\Delta m + A)/\sqrt{2\sigma_{\rm lc}^2}\bigg]}{1 + {\rm erf}\bigg[(-\Delta m + A)/\sqrt{2\sigma_{\rm lc}^2} \bigg]}\ ,
\end{eqnarray}
where ${\rm erf}(x)$ is the Gauss error function. 

When the sample with fixed $m^\prime$ is frozen at $t_0$ and re-observed at late times (or at earlier times), the mean flux will be fainter by $|A|$ magnitude, and there will be an overabundance of dimmed quasars expected from Eqn.~(\ref{eqn:toy}). The probability distribution of magnitude will be a Gaussian with mean $m^\prime - A$ and dispersion $\sigma_{\rm lc}$. If the time lapse since sample selection is longer than the characteristic damping timescale, the RMS variation $\sigma_{\rm lc}\approx {\rm SF}_{\infty}/\sqrt{2}$, where ${\rm SF}_{\infty}$ is the asymptotic structure function on very long timescales. Given a typical bright-end slope of the LF $\gamma=0.8$ and $\sigma_{\rm lc}=0.2$, we have $A\sim -0.1$ mag, which is a small effect to measure with a large sample of quasars \citep{Caplar_etal_2020}, and an overabundance of dimmed quasars of $N_{\rm dimmed}/N_{\rm brightened}\approx 1.8, 3.5$ for $|\Delta m|=0, 0.25$. The comparisons above are between the epoch of sample selection and a different epoch. If we move sufficiently far away from the selection epoch (typically a few years, e.g., the damping timescale multiplied by $1+z$), the distribution of observed magnitude will converge to the same Gaussian with mean $m^\prime - A$ and dispersion $\sigma_{\rm lc}$, and the time asymmetry between two different epochs will largely go away. 

In practice, the real flux-limited sample covers a range of magnitudes, redshifts and variability characteristics, and the effects of this sample bias are best studied with simulated data (\S\ref{sec:sims}). Nevertheless, this toy model provides an intuitive understanding of the observed trends. 

\subsection{Simulated quasar sample and light curves}\label{sec:sims}

We start by generating a large sample of quasars at $0.3<z<5$ following the observed optical luminosity function (LF) at different redshifts \citep{Hopkins_etal_2007}. The lower redshift cut is to reduce the impact of host galaxy light on quasar variability. The mock quasar sample was generated to a limiting magnitude of $i_{\rm mean}=22$, where $i_{\rm mean}$ is the mean magnitude of the quasar. We chose this limiting magnitude above which the quasar LF is well measured from large spectroscopic surveys \citep[e.g.,][]{Richards_etal_2006a}. The imposed flux limits to define our quasar sample for variability studies will be far from $i=22$, therefore missing the $i>22$ faint quasar population will not affect our results. Fig.~\ref{fig:Lz} (top) shows the distribution of simulated quasars in the $L-z$ plane. There are in total $\sim 200,000$ simulated quasars at $i<22$, ensuring we have sufficient statistics when restricting the sample to brighter flux limits. 

For each simulated quasar, we assign luminosity and black hole mass following empirical relations derived for SDSS quasars \citep{Richards_etal_2006a,Shen_etal_2011}. We then use the empirical relations in \citet{MacLeod_etal_2010} to assign DRW parameters and generate a stochastic $i$-band light curve for each quasar with a daily cadence over an observed baseline of 20 years. The mock light curve varies around the mean $i$-band magnitude of the quasar. 

For each light curve, we compute the rest-frame SF: 
\begin{equation}
{\rm SF}(\tau)=\bigg[ \frac{1}{N(\tau)}\sum_{i<j}(x_i-x_j)^2\bigg]^{1/2}\ ,
\end{equation}
where $i$ and $j$ are indices of the time series, $N(\tau)$ is the number of pairs in each rest-frame time $\tau$ bin of the SF, and $x_i$ and $x_j$ are the corresponding magnitudes at times $t_i$ and $t_j$ for these pairs. ${\rm SF_{ic}}$ and ${\rm SF_{dc}}$ are computed using pairs with increasing and decreasing fluxes, respectively; ${\rm SF_{tot}}$ uses all pairs. We verify that the ensemble SF is consistent with the DRW prediction, while individual light curves (particularly high-redshift quasars with long observed-frame damping timescales) may deviate from the DRW prediction due to stochasticity over the limited observing baseline. 

To define a flux-limited sample constructed at an earlier epoch, we use the first light curve point at $t_0$. This is a key step to properly mimic the real sample selection. In earlier studies \citep[e.g.,][]{Tachibana_etal_2020}, while not explicitly stated, we suspect that the mock sample was constructed using the mean magnitude instead of the instantaneous magnitude at the selection epoch. This detail will lead to distinctive outcomes from simulated light curves, as we demonstrate below. 

\begin{figure}
  \centering
    \includegraphics[width=0.48\textwidth]{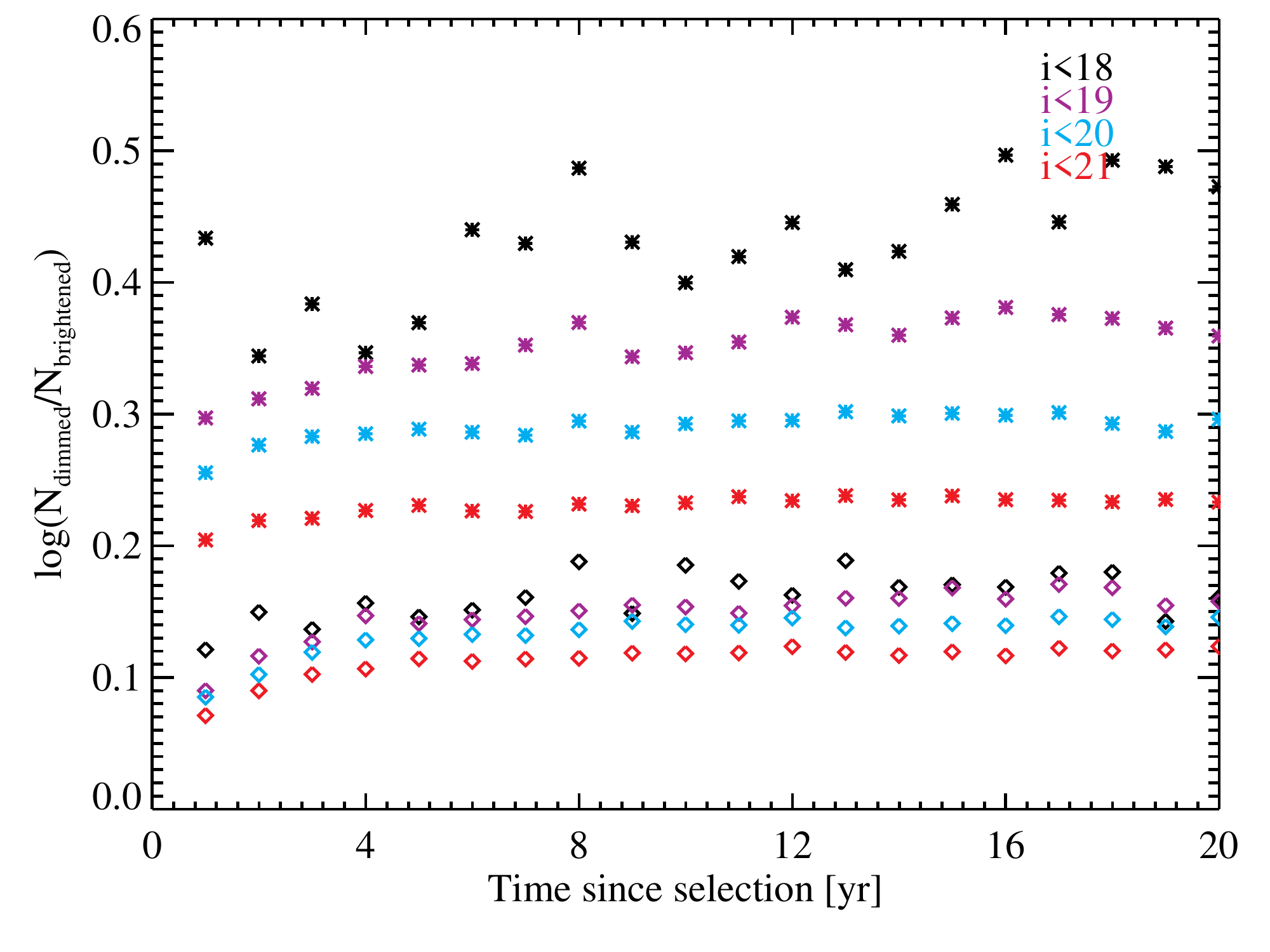}
    \caption{Ratio of the numbers of dimmed and brightened quasars as a function of time. Different colors denote different flux limits in the sample. The diamonds are for a magnitude change threshold of $|\Delta m|=0$ and the asterisks are for a magnitude change threshold of $|\Delta m|=0.25$.  }
    \label{fig:mag_bias}
\end{figure}

\citet{Rumbaugh_etal_2018} and \citet{Luo_etal_2020} have shown that when observing a flux-limited quasar sample in a later survey, dimmed quasars will outnumber brightened quasars and this asymmetry increases when the threshold of the magnitude change $|\Delta m|$ increases. While the exact levels of the asymmetry depend on the simulated quasar population (e.g., the shape of the LF)\footnote{Our adopted LF has a power-law form at the bright end; a steeper drop of the bright-end LF would lead to more severe bias for brighter samples.} and adopted DRW parameters, we confirm these general trends in Fig.~\ref{fig:mag_bias} using our simulated quasar samples at various flux limits and thresholds in magnitude changes. 

While the asymmetry in the numbers of dimmed and brightened quasars is prominent, the mean magnitude change of the whole sample is mild, since most of the time pairs have small differences in magnitude. The mean magnitude differences after 10 years in the observed frame for our simulated sample of $i_{t_0}<19$ is $\sim 0.1$~mag fainter, roughly matching the observed mean magnitude change in \citet{Caplar_etal_2020}. This mean magnitude bias is stronger at lower redshifts (at fixed flux limit), since the RMS variability amplitude is higher (Fig.~\ref{fig:Lz} and Eqn.~\ref{eqn:mag_bias}), again consistent with the findings in \citet{Caplar_etal_2020}.  

\begin{figure}
  \centering
    \includegraphics[width=0.48\textwidth]{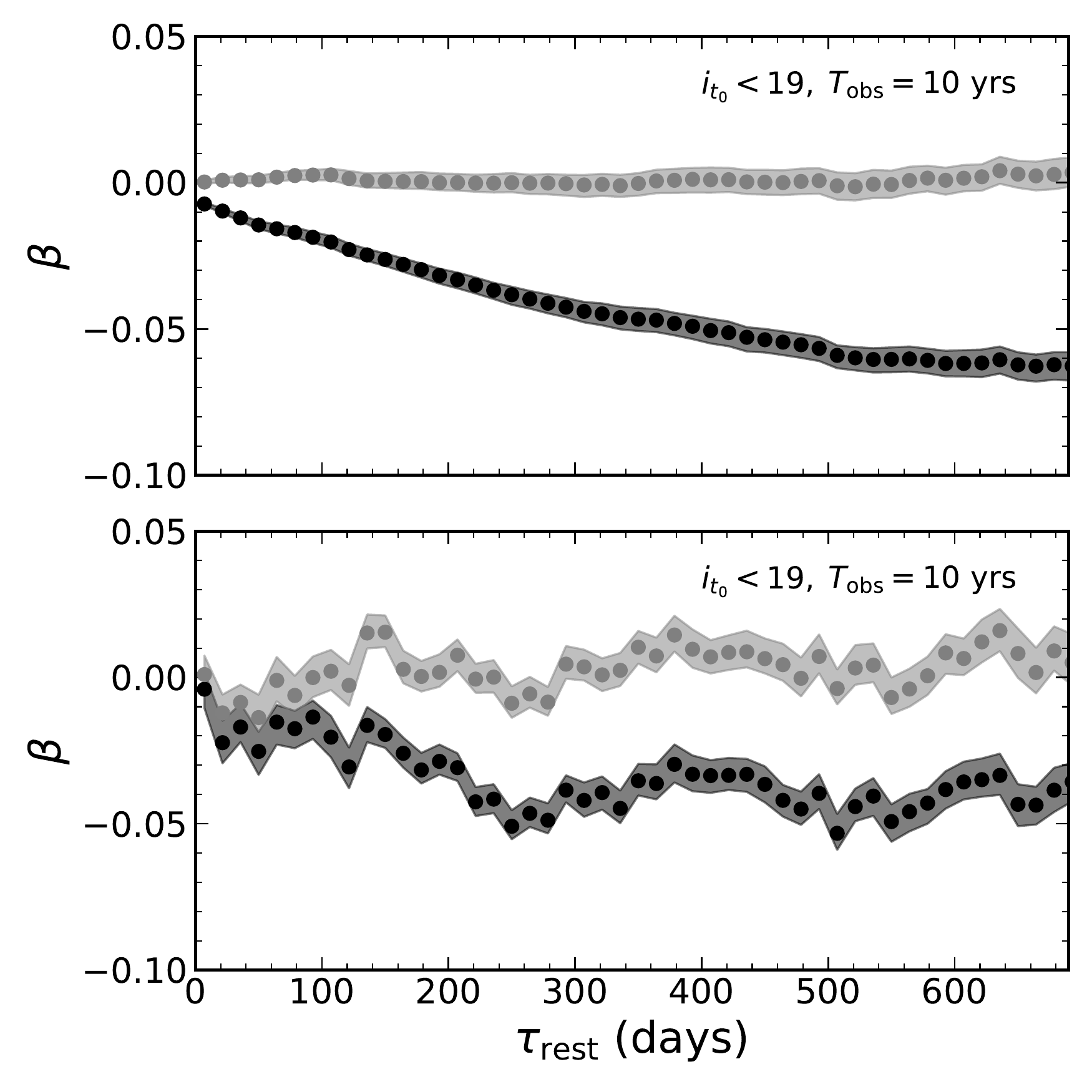}
    \caption{Time asymmetry in the SF for a simulated quasar sample of $i_{t_0}<19$ over a 10-yr baseline in the observed frame. {\em Top:} results from the ideal light curves. {\em Bottom:} results from downgraded light curves that mimic the CRTS light curves. In each panel, the black dots and shaded band are the measured SF asymmetry $\beta(\tau)$ and $1\sigma$ uncertainties for the flux-limited sample, and the gray dots and shaded band are the results for a comparison sample with $i_{\rm mean}<19$; the latter is not a flux-limited sample at selection. Uncertainties on $\beta$ are estimated from bootstrap resampling. }
    \label{fig:beta}
\end{figure}

Fig.~\ref{fig:beta} shows the resulting asymmetry in the ensemble SF measurements, using a fiducial simulated quasar sample with $i_{t_0}<19$ and a 10-yr observational baseline that starts at the selection epoch $t_0$. This flux-limited mock sample and light curve duration roughly match those of the bright CRTS quasars studied in \citet{Tachibana_etal_2020}. Since the CRTS sample is a bright quasar sample ($V<18$), most of the quasars come from SDSS-I/II legacy surveys based on SDSS targeting photometry a few years earlier than the bulk of the CRTS light curves. 

The top panel of Fig.~\ref{fig:beta} shows the results based on the noise-free, daily-sampled mock light curves, and the bottom panel shows the results based on downgraded light curves to mimic the cadence, seasonal gaps and magnitude uncertainties in the CRTS sample. The results from downgraded light curves are noisier than those from the ideal light curves, but still reproduce the overall trend of $\beta$.  
As expected, the flux-limited sample defined at $t_0$ results in an asymmetric SF such that ${\rm SF_{ic}}$ is systematically smaller at the few percent level than ${\rm SF_{dc}}$, similar to the observed trend in the CRTS sample. For comparison, the gray lines in Fig.~\ref{fig:beta} show the results of a control sample selected with $i_{\rm mean}<19$, where $\beta$ is consistent with zero as expected. Uncertainties are estimated based on bootstrap resampling.  


In our fiducial calculations of $\beta$, we have used the mean SFs over the sample. Using the median SFs, or the median/mean of the $\beta$ distribution from individual quasars produces similar results. 

\section{Discussion}\label{sec:disc}


\begin{figure*}
  \centering
    \includegraphics[width=0.45\textwidth]{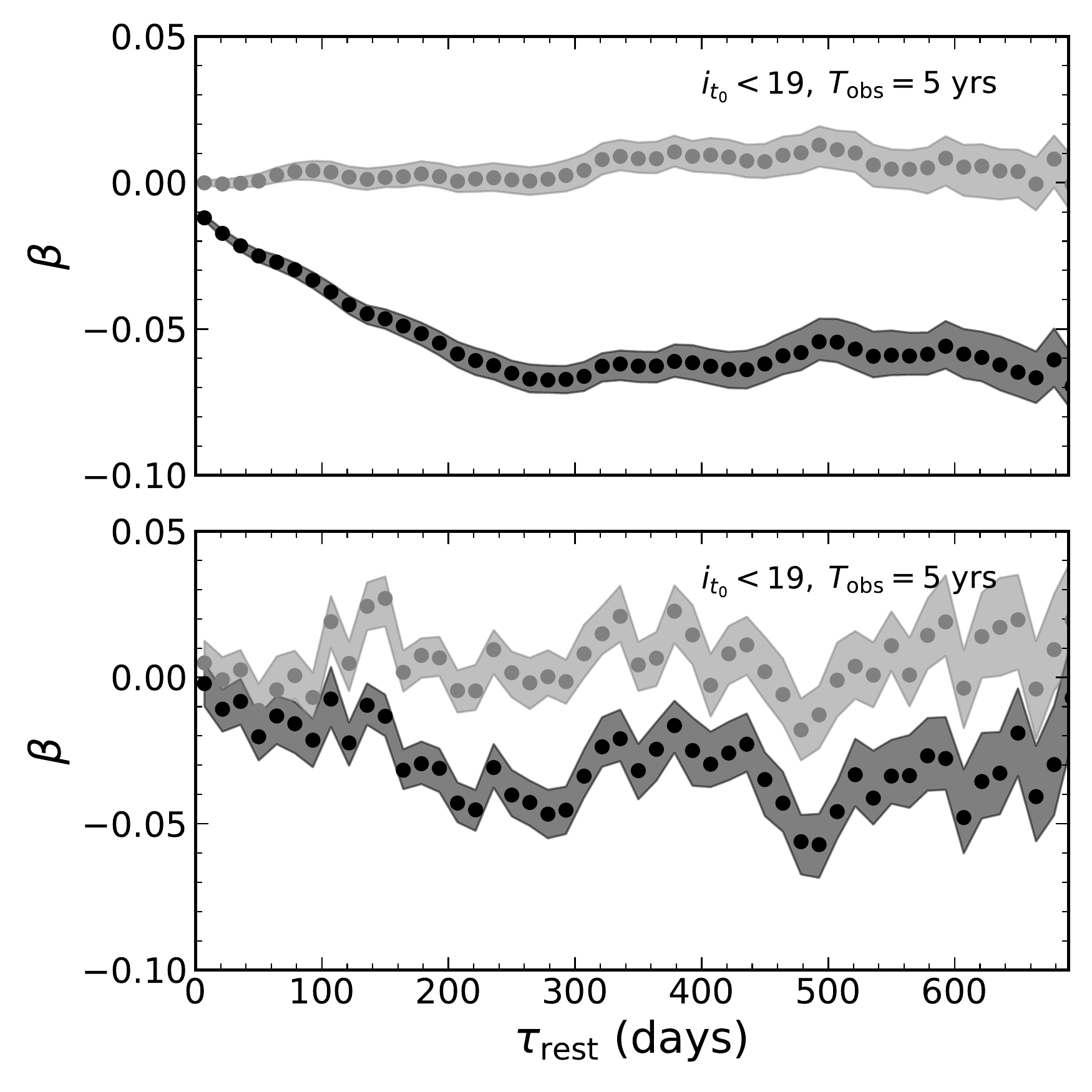}
    \includegraphics[width=0.45\textwidth]{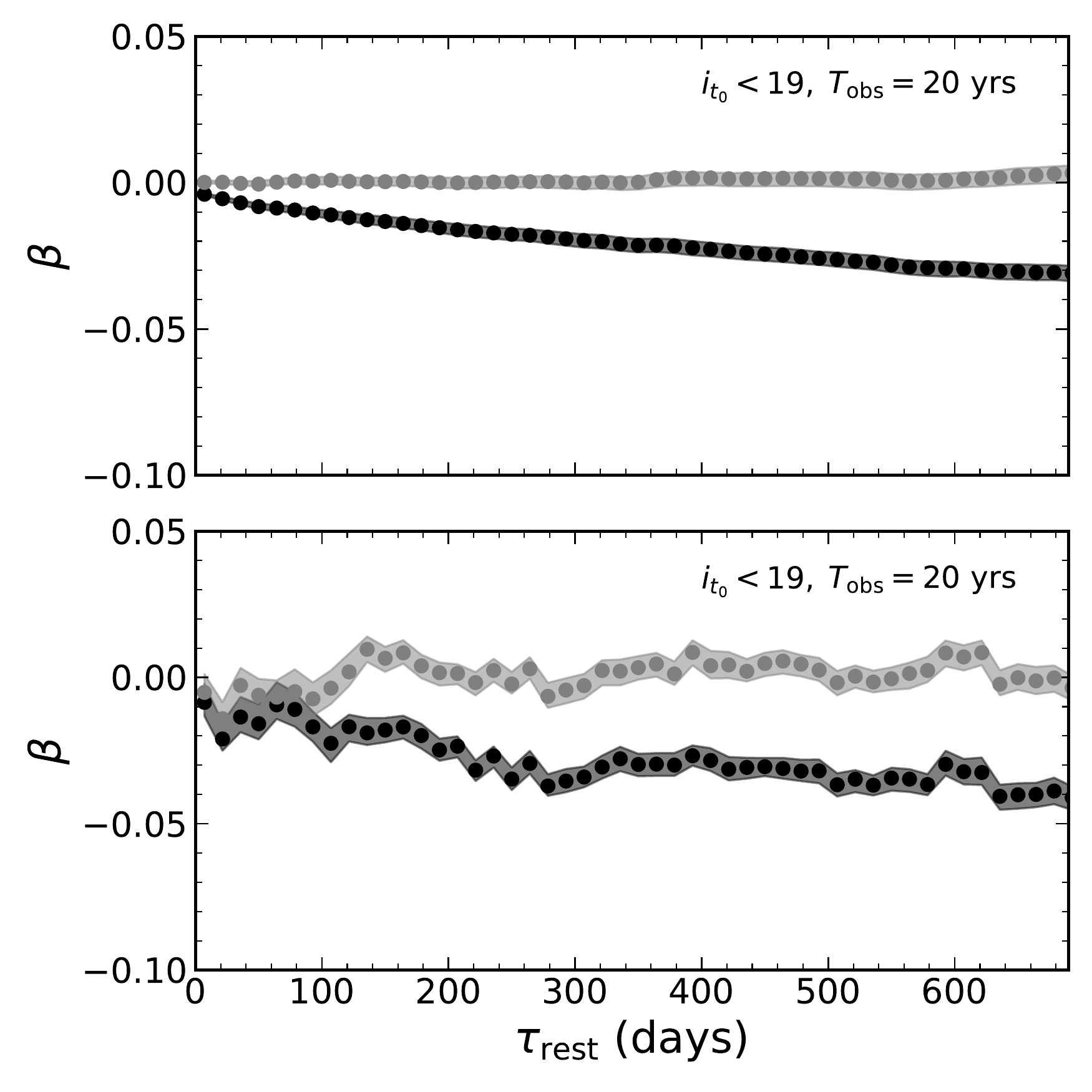}
    \caption{Similar to Fig.~\ref{fig:beta}, but for different light curve durations. SF measurements are noisier for shorter durations due to fewer pairs from the light curve. }
    \label{fig:beta_duration}
\end{figure*}

\begin{figure*}
  \centering
    \includegraphics[width=0.45\textwidth]{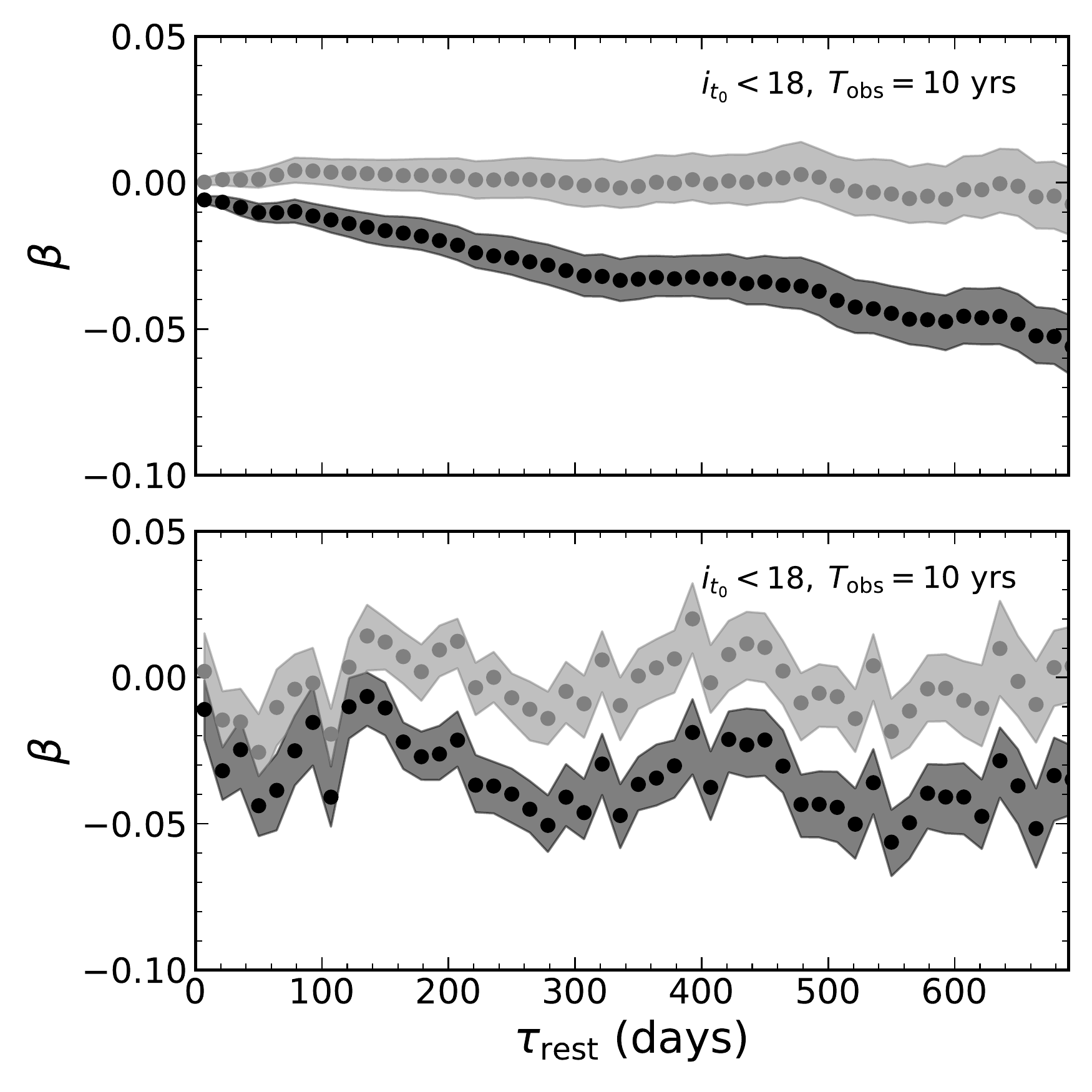}
    \includegraphics[width=0.45\textwidth]{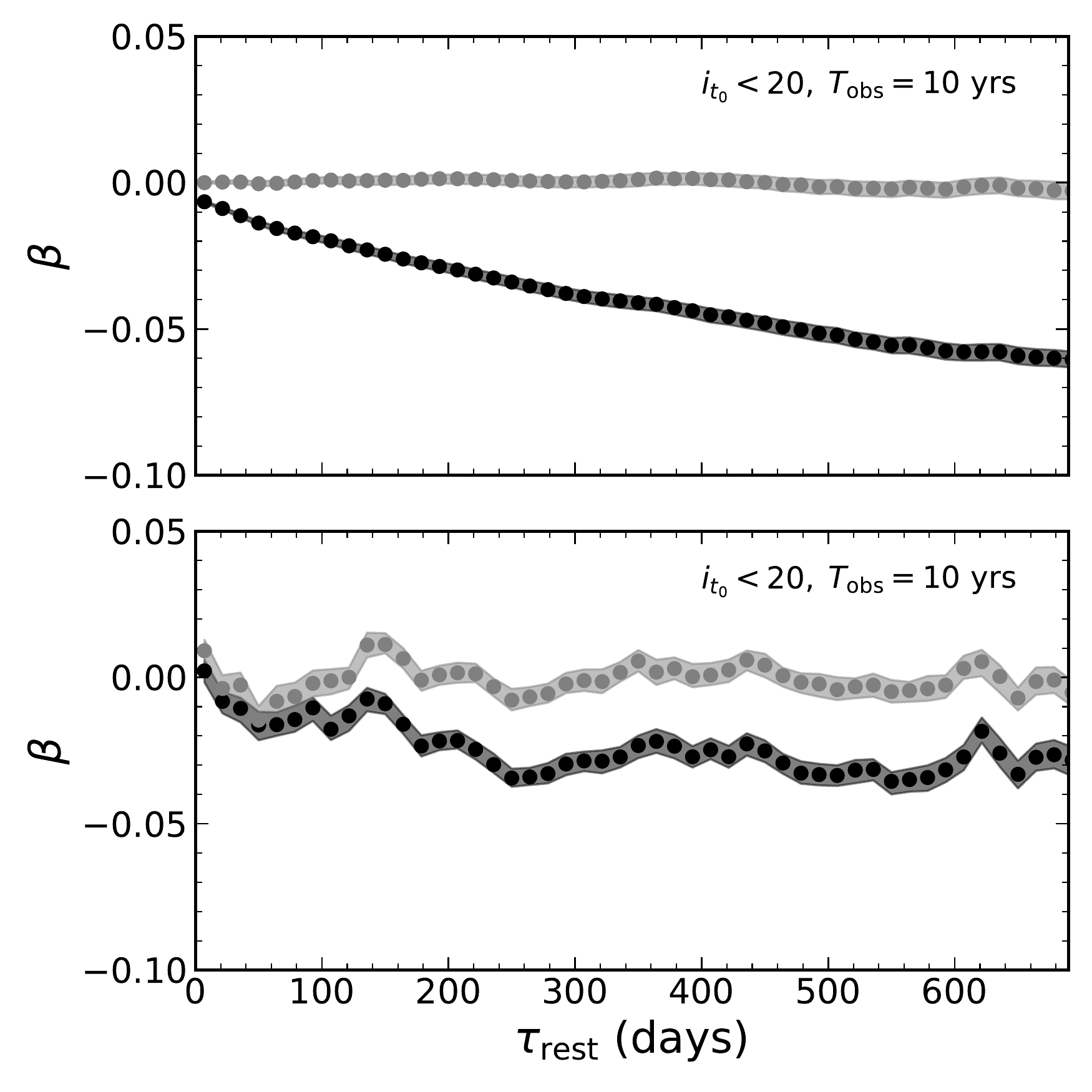}
    \caption{Similar to Fig.~\ref{fig:beta}, but for different flux limits at the selection. SF measurements are noisier for brighter limits due to fewer quasars in the sample. }
    \label{fig:beta_ilim}
\end{figure*}

Figs.~\ref{fig:beta_duration} and \ref{fig:beta_ilim} show the resulting $\beta$ for different flux limits and light curve lengths. The time asymmetry of the SF is insensitive to the flux limit, mainly due to the combined effects of LF steepening and decreasing variability towards the bright end (see Eqn.~\ref{eqn:mag_bias} and Fig.~\ref{fig:Lz}). On the other hand, this bias is somewhat alleviated for longer baselines. The latter trend is straightforward to understand (\S\ref{sec:toy}). Pairs of epochs contributing to the SF that are both far away from the selection epoch do not retain the memory of the time asymmetry. The longer the length of the light curve, the more dilution from these epochs to reduce the overall time asymmetry caused by pairs that include the selection epoch. In addition, the decorrelation timescale (i.e., the DRW damping timescale) will also affect the exact level of this asymmetry in the ensemble SF. Our adopted DRW damping timescales from \citet{MacLeod_etal_2010} are reasonable, but some quasars may have even longer damping timescales to worsen this sample bias. 


Overall, our simulations predict negative $\beta$ over multi-year baselines roughly starting from the selection epoch. \citet{Tachibana_etal_2020} also found positive $\beta$ for rest-frame SF time $\tau\lesssim100$\,days (peaking around $\sim 40$\,days). The SF measurements are more affected by magnitude uncertainties at shorter timescales due to the smaller SF amplitude. In addition, some of these short-timescale pairs could come from CRTS photometry prior to the SDSS selection that captured more rising segments of the light curve. Nevertheless, it is possible that this positive $\beta$ at $\tau\lesssim100$\,days is real, and may reflect underlying driving mechanisms of the short-time variability \citep[e.g.,][]{Kawaguchi_etal_1998,Tachibana_etal_2020}. 

This sample bias also explains negative and zero $\beta$ values reported in other studies of quasar light curves. If the SFs are computed using historic photometric data for a flux-limited sample defined at a later epoch, then this bias will lead to on average positive $\beta$ values, as reported in \citet{deVries_etal_2005} with historic imaging data for SDSS quasars (the photometry at the SDSS epoch was also used in the SF calculation). However, if the light curves are far away from the sample selection epoch, or if the sample is not a flux-limited sample or there is no clearly-defined selection epoch, then there should be no time asymmetry in the ensemble SF \citep[$\beta\approx 0$, e.g.,][]{Hawkins_2002,Chen_Wang_2015}. We have confirmed these predicted trends of $\beta$ with our simulated data.

\section{Conclusions}\label{sec:con}

With ever increasing sample statistics and extended baselines for variability studies, it has now become feasible to explore subtle ensemble variability characteristics of quasars. It is notoriously difficult to interpret light curves with physical models to understand the nature of quasar variability, given the broad range of stochastic processes that can describe the observed time series \citep[e.g.,][]{Scargle_2020}. Salient ideas have been proposed, which use simple statistics of the time series to explore specific aspects of quasar variability, e.g., the time-asymmetry in the ensemble SF \citep[e.g.,][]{Kawaguchi_etal_1998}, the rms-flux relation \citep{Uttley_McHardy_2001}, and characteristic timescales in the variability power spectrum \citep[e.g.,][]{McHardy_etal_2006,Burke_etal_2021}. The hope is that with the correct interpretation of these statistics, one can infer the physical mechanisms of quasar variability. 

In this work we highlighted a sample bias introduced at the time of sample selection, in light of recent reports of time asymmetry in optical quasar variability observed over extended periods. With simulated quasars that follow the observed luminosity-redshift distribution, and simulated time-symmetric light curves based on the DRW model, we demonstrated the effects of this bias for flux-limited samples selected at an earlier epoch and re-observed at later times with a multi-year baseline: 

\begin{enumerate}
    \item there will be more dimmed than brightened quasars observed at later times, and the sample mean flux will fade by a small amount ($\sim 0.1$\,mag); 
    \item there is an asymmetry in the SF measured from increasing-flux pairs and decreasing-flux pairs at the few percent level.
\end{enumerate}

Both predictions based on simulated data are similar to observed trends \citep{Voevodkin_2011,Caplar_etal_2020,Tachibana_etal_2020}. The selection of a flux-limited quasar sample at certain epoch inevitably incorporates an asymmetry in variability, since there are more temporarily brightened quasars above the flux limit than those falling below. This Eddington-like bias is common in statistical studies of survey samples. Depending on what light curves are used (i.e., before or after the sample selection epoch, and the length of the light curves), the $\beta$ statistic defined in Eqn.~(\ref{eqn:beta}) can either be positive, negative or zero (\S\ref{sec:disc}).

While our results do not necessarily imply that the DRW model is correct, they provide a cautionary note on recent claims that there is a time asymmetry in quasar light curves based on bright, (approximately) flux-limited samples. The manifestation of this bias, in terms of the mean magnitude changes and the asymmetry in the SF, is generally small and only detectable with large statistical quasar samples as those studied in, e.g., \citet{Caplar_etal_2020} and \citet{Tachibana_etal_2020}.  

To mitigate this bias, one could potentially do one of the following: (1) use the later-epoch photometry to select a second flux-limited sample (with the same flux limit) that will recover those missed in the initial sample, and average the results from the two flux-limited samples; (2) carefully adjust the initial sample to include equal numbers of brightened and faded quasars in each $|\Delta {\rm mag}|$ bin; or (3) use portions of light curves that are at least a few years away from the sample selection epoch. On the other hand, if the quasar LF can be accurately measured, the observed apparent asymmetry in ensemble quasar light curves may be used to provide independent constraints on quasar variability, e.g., the luminosity dependences of the variability amplitude and damping (decorrelation) timescale. 

\acknowledgments

We thank the referee for useful comments that improved the presentation of this work. Y.S. was supported by NSF grant AST-2009947. C.J.B. acknowledges support from the Illinois Graduate Survey Science Fellowship.

\bibliography{lc_comp_rv1.bbl}

\begin{thebibliography}{}
\expandafter\ifx\csname natexlab\endcsname\relax\def\natexlab#1{#1}\fi
\providecommand{\url}[1]{\href{#1}{#1}}

\bibitem[{{Bauer} {et~al.}(2009){Bauer}, {Baltay}, {Coppi}, {Ellman}, {Jerke},
  {Rabinowitz}, \& {Scalzo}}]{Bauer_etal_2009}
{Bauer}, A., {Baltay}, C., {Coppi}, P., {et~al.} 2009, \apj, 696, 1241

\bibitem[{{Burke} {et~al.}(2021)}]{Burke_etal_2021}
{Burke}, C.~J, {Shen}, Y., {Blaes}, O., {et~al.} 2021, Science, in press

\bibitem[{{Caplar} {et~al.}(2017){Caplar}, {Lilly}, \&
  {Trakhtenbrot}}]{Caplar_etal_2017}
{Caplar}, N., {Lilly}, S.~J., \& {Trakhtenbrot}, B. 2017, \apj, 834, 111

\bibitem[{{Caplar} {et~al.}(2020){Caplar}, {Pena}, {Johnson}, \&
  {Greene}}]{Caplar_etal_2020}
{Caplar}, N., {Pena}, T., {Johnson}, S.~D., \& {Greene}, J.~E. 2020, \apjl,
  889, L29

\bibitem[{{Chen} \& {Wang}(2015)}]{Chen_Wang_2015}
{Chen}, X.-Y., \& {Wang}, J.-X. 2015, \apj, 805, 80

\bibitem[{{de Vries} {et~al.}(2005){de Vries}, {Becker}, {White}, \&
  {Loomis}}]{deVries_etal_2005}
{de Vries}, W.~H., {Becker}, R.~H., {White}, R.~L., \& {Loomis}, C. 2005, \aj,
  129, 615

\bibitem[{{Eddington}(1913)}]{Eddington_1913}
{Eddington}, A.~S. 1913, \mnras, 73, 359

\bibitem[{{Giveon} {et~al.}(1999){Giveon}, {Maoz}, {Kaspi}, {Netzer}, \&
  {Smith}}]{Giveon_etal_1999}
{Giveon}, U., {Maoz}, D., {Kaspi}, S., {Netzer}, H., \& {Smith}, P.~S. 1999,
  \mnras, 306, 637

\bibitem[{{Hawkins}(2002)}]{Hawkins_2002}
{Hawkins}, M.~R.~S. 2002, \mnras, 329, 76

\bibitem[{{Hopkins} {et~al.}(2007){Hopkins}, {Richards}, \&
  {Hernquist}}]{Hopkins_etal_2007}
{Hopkins}, P.~F., {Richards}, G.~T., \& {Hernquist}, L. 2007, \apj, 654, 731

\bibitem[{{Kawaguchi} {et~al.}(1998){Kawaguchi}, {Mineshige}, {Umemura}, \&
  {Turner}}]{Kawaguchi_etal_1998}
{Kawaguchi}, T., {Mineshige}, S., {Umemura}, M., \& {Turner}, E.~L. 1998, \apj,
  504, 671

\bibitem[{{Kelly} {et~al.}(2009){Kelly}, {Bechtold}, \&
  {Siemiginowska}}]{Kelly_etal_2009}
{Kelly}, B.~C., {Bechtold}, J., \& {Siemiginowska}, A. 2009, \apj, 698, 895

\bibitem[{{Luo} {et~al.}(2020){Luo}, {Shen}, \& {Yang}}]{Luo_etal_2020}
{Luo}, Y., {Shen}, Y., \& {Yang}, Q. 2020, \mnras, 494, 3686

\bibitem[{{MacLeod} {et~al.}(2010){MacLeod}, {Ivezi{\'c}}, {Kochanek},
  {Koz{\l}owski}, {Kelly}, {Bullock}, {Kimball}, {Sesar}, {Westman}, {Brooks},
  {Gibson}, {Becker}, \& {de Vries}}]{MacLeod_etal_2010}
{MacLeod}, C.~L., {Ivezi{\'c}}, {\v Z}., {Kochanek}, C.~S., {et~al.} 2010,
  \apj, 721, 1014

\bibitem[{{McHardy} {et~al.}(2006){McHardy}, {Koerding}, {Knigge}, {Uttley}, \&
  {Fender}}]{McHardy_etal_2006}
{McHardy}, I.~M., {Koerding}, E., {Knigge}, C., {Uttley}, P., \& {Fender},
  R.~P. 2006, \nat, 444, 730

\bibitem[{{Mushotzky} {et~al.}(2011){Mushotzky}, {Edelson}, {Baumgartner}, \&
  {Gandhi}}]{Mushotzky_etal_2011}
{Mushotzky}, R.~F., {Edelson}, R., {Baumgartner}, W., \& {Gandhi}, P. 2011,
  \apjl, 743, L12

\bibitem[{{Richards} {et~al.}(2006){Richards}, {Strauss}, {Fan}, {Hall},
  {Jester}, {Schneider}, {Vanden Berk}, {Stoughton}, {Anderson}, {Brunner},
  {Gray}, {Gunn}, {Ivezi{\'c}}, {Kirkland}, {Knapp}, {Loveday}, {Meiksin},
  {Pope}, {Szalay}, {Thakar}, {Yanny}, {York}, {Barentine}, {Brewington},
  {Brinkmann}, {Fukugita}, {Harvanek}, {Kent}, {Kleinman}, {Krzesi{\'n}ski},
  {Long}, {Lupton}, {Nash}, {Neilsen}, {Nitta}, {Schlegel}, \&
  {Snedden}}]{Richards_etal_2006a}
{Richards}, G.~T., {Strauss}, M.~A., {Fan}, X., {et~al.} 2006, \aj, 131, 2766

\bibitem[{{Rumbaugh} {et~al.}(2018){Rumbaugh}, {Shen}, {Morganson}, {Liu},
  {Banerji}, {McMahon}, {Abdalla}, {Benoit-L{\'e}vy}, {Bertin}, {Brooks},
  {Buckley-Geer}, {Capozzi}, {Carnero Rosell}, {Carrasco Kind}, {Carretero},
  {Cunha}, {DAndrea}, {da Costa}, {DePoy}, {Desai}, {Doel}, {Frieman},
  {Garc{\'{\i}}a-Bellido}, {Gruen}, {Gruendl}, {Gschwend}, {Gutierrez},
  {Honscheid}, {James}, {Kuehn}, {Kuhlmann}, {Kuropatkin}, {Lima}, {Maia},
  {Marshall}, {Martini}, {Menanteau}, {Plazas}, {Reil}, {Roodman}, {Sanchez},
  {Scarpine}, {Schindler}, {Schubnell}, {Sheldon}, {Smith}, {Soares-Santos},
  {Sobreira}, {Suchyta}, {Swanson}, {Walker}, {Wester}, \& {(DES
  Collaboration}}]{Rumbaugh_etal_2018}
{Rumbaugh}, N., {Shen}, Y., {Morganson}, E., {et~al.} 2018, \apj, 854, 160

\bibitem[{{Scargle}(2020)}]{Scargle_2020}
{Scargle}, J.~D. 2020, \apj, 895, 90

\bibitem[{{Shen} \& {Kelly}(2010)}]{Shen_Kelly_2010}
{Shen}, Y., \& {Kelly}, B.~C. 2010, \apj, 713, 41

\bibitem[{{Shen} {et~al.}(2011){Shen}, {Richards}, {Strauss}, {Hall},
  {Schneider}, {Snedden}, {Bizyaev}, {Brewington}, {Malanushenko},
  {Malanushenko}, {Oravetz}, {Pan}, \& {Simmons}}]{Shen_etal_2011}
{Shen}, Y., {Richards}, G.~T., {Strauss}, M.~A., {et~al.} 2011, \apjs, 194, 45

\bibitem[{{Tachibana} {et~al.}(2020){Tachibana}, {Graham}, {Kawai},
  {Djorgovski}, {Drake}, {Mahabal}, \& {Stern}}]{Tachibana_etal_2020}
{Tachibana}, Y., {Graham}, M.~J., {Kawai}, N., {et~al.} 2020, \apj, 903, 54

\bibitem[{{Uttley} \& {McHardy}(2001)}]{Uttley_McHardy_2001}
{Uttley}, P., \& {McHardy}, I.~M. 2001, \mnras, 323, L26

\bibitem[{{Vanden Berk} {et~al.}(2004){Vanden Berk}, {Wilhite}, {Kron},
  {Anderson}, {Brunner}, {Hall}, {Ivezi{\'c}}, {Richards}, {Schneider}, {York},
  {Brinkmann}, {Lamb}, {Nichol}, \& {Schlegel}}]{VandenBerk_etal_2004}
{Vanden Berk}, D.~E., {Wilhite}, B.~C., {Kron}, R.~G., {et~al.} 2004, \apj,
  601, 692

\bibitem[{{Voevodkin}(2011)}]{Voevodkin_2011}
{Voevodkin}, A. 2011, arXiv e-prints, arXiv:1107.4244

\end{thebibliography}

\end{document}